\documentclass{article}
\usepackage{epsfig}
\usepackage{amsmath}
\usepackage{amssymb}
\usepackage[hang,nooneline,scriptsize]{subfigure}

\usepackage[dvips]{color}
\usepackage[normalem]{ulem}

\newcommand{\ee}{\end{equation}}
\newcommand{\eea}{\end{eqnarray}}
\newcommand{\be}{\begin{equation}}
\newcommand{\bea}{\begin{eqnarray}}

\begin{document}

\title{\LARGE \bf Charged Boson Stars and Black Holes with Non-Minimal Coupling to Gravity}
  \author{
  \large  Y. Verbin$^a$  $\:${\em and}$\:$
  Y. Brihaye$^b$ \thanks{Electronic addresses: verbin@openu.ac.il ; yves.brihaye@umons.ac.be  } }
 \date{ }
   \maketitle
    \centerline{$^a$ \em Department of Natural Sciences, The Open University
   of Israel,}
   \centerline{\em Raanana 43107, Israel}
     \vskip 0.4cm
    \centerline{$^b$ \em Physique Th\'eorique et Math\'ematiques, Universit\'e de Mons,}
   \centerline{\em Place du Parc, B-7000  Mons, Belgique}

 \maketitle
\begin{abstract}
We find new spherically symmetric charged boson star solutions of a complex scalar field coupled non-minimally to gravity by a ``John-type'' term of Horndeski theory, that is a coupling between the kinetic scalar term and Einstein tensor. We study the parameter space of the solutions and find two distinct families according to their position in parameter space. More widespread is the family of  solutions (which we call branch 1) existing for a finite interval of the central value of the scalar field starting from zero and ending at some finite maximal value. This branch contains
as a special case the charged boson stars of the minimally coupled theory. In some regions of parameter space we find a new second branch (``branch 2'') of solutions which are more massive and more stable than those of branch 1. This second branch exists also in a finite interval of the central value of the scalar field, but its end points (either both or in some cases only one) are extremal Reissner-Nordstr\"{o}m black hole solutions.
\end{abstract}

\maketitle


\section{Introduction}\label{Introduction}
\setcounter{equation}{0}

Following the resurrection \cite{DeffayetEtAl2011,KobayashiEtAl2011,CharmousisEtAl2012a,deRham-Heisenberg,CharmousisEtAl2012b}
 of Horndeski scalar-tensor theory \cite{Horndeski1974} as an early realization of Galileon gravity  \cite{NicolisEtAl2008}, a large number of investigations have been carried out in numerous aspects of the theory, from cosmology and large scale structure of the universe to small scale of astrophysical compact objects like boson stars and black holes.

Horndeski theory has a huge freedom encoded in four arbitrary functions of the scalar field $\phi$ and its standard kinetic term $(\partial \phi )^2$ -- see e.g. \cite{KobayashiEtAl2011}. It is therefore very difficult to analyze without specifying somehow these four functions. An interesting special subclass of the theory, known as the ``Fab Four'' \cite{CharmousisEtAl2012a,CharmousisEtAl2012b} is selected by the {\it screening} or {\it self-tuning} property which means that even when a ``bare'' cosmological constant is present in the Lagrangian, the theory admits Minkowski spacetime as a solution. Although the ``Fab Four'' is much simpler, its Lagrangian is still a sum of four terms
\be \label{F4}
{\cal L} = V_{J}(\phi)G^{\mu \nu} \partial_{\mu}\phi \partial_{\nu} \phi +
V_{P} (\phi) ^{**}R^{\kappa \lambda \mu \nu} (\nabla_\kappa \phi )(\nabla_\mu \phi )\nabla_\lambda \nabla_\nu \phi + V_{G} (\phi) R + V_{R} (\phi) \Re_{GB}
,\ee
with four arbitrary functions. $\Re_{GB}$  is the Gauss-Bonnet term and $^{**}R^{\kappa \lambda \mu \nu}$ is the double-dual of Riemann tensor.

Usually, further simplifications are made like the Sotiriou-Zhou model  \cite{Sotiriou-Zhou2013,Sotiriou-Zhou2014} who took $V_{J}(\phi)=V_{P}(\phi)=0$, $V_{G}(\phi)=1/(16\pi \cal {G})$, $V_{R}(\phi)$ proportional to $\phi$ and added the standard scalar kinetic term, or the so-called ``Fab Five'' \cite{ApplebyEtAl2012,StarobinskyEtAl2016} which makes a different choice of $V_{P}(\phi)=V_{R}(\phi)=0$, $V_{G}(\phi)$ standard Einstein-Hilbert as before and $V_{J}(\phi)=-\beta$, a constant. The new ingredient in this theory is the non-minimal coupling term of
the scalar kinetic term to the Einstein tensor, ${\cal L}_{NM} \sim G^{\mu \nu} \partial_{\mu}\phi \partial_{\nu} \phi$ -- the so-called ``John'' contribution in the terminology of Charmousis {\it et al.} \cite{CharmousisEtAl2012b}.

Quite recently, black hole solutions\cite{Rinaldi2012,Babichev+Charmousis2014,AnabalonEtAl2013,Minamitsuji2014} were discovered in this ``John''-type version of Horndeski theory. Black hole solutions \cite{CharmousisEtAl2014} were found also in a ``biscalar extension'' of the theory and boson stars were found \cite{BrihayeEtAl2016} in a different biscalar version where the scalar field becomes complex. On the other hand, the real scalar system was also studied in the presence of a radial electric field \cite{KolyvarisEtAl2011,Cisterna-Erices2014} and black holes were found in that case too. These latter black hole solutions are essentially a ``superposition'' of the pure scalar black holes with the electric field of a point charge at their centers. Further study \cite{BabichevEtAl2015} of the real scalar system produced magnetic black holes as well.

It is evident from the above description that still missing  is a study of the gauged version of the ``John'' Lagrangian mentioned above thus producing the conventional minimal coupling between
the scalar and vector fields. The dynamics of such a system is determined by the following action (we stay Abelian):
\be \label{totalAction}
S = \int d^4 x \sqrt{- g} \left (\frac{R}{2 \kappa} -\frac{1}{4} F_{\mu\nu}F^{\mu\nu}+ \frac{1}{2}D_{\mu}\Phi^* D^{\mu} \Phi - U(|\Phi |) -\beta G^{\mu \nu} D_{\mu}\Phi^* D_{\nu} \Phi \right )
\ee
where  $D_{\mu}\Phi = \partial_{\mu}\Phi - ie A_{\mu}\Phi $ is the gauge covariant derivative of the scalar complex field $\Phi$  with respect to the vector potential and $F_{\mu\nu}=\partial_{\mu}A_{\nu}-\partial_{\nu}A_{\mu}$ is the corresponding field strength. $R$ is Ricci scalar, $G^{\mu \nu}$ is the Einstein tensor and $\kappa$ is given in terms of  Newton's constant $\cal {G}$ by $\kappa=8 \pi \cal {G}$ or in terms of the Planck mass (in the appropriate units) as $\kappa=1/m^2_{Pl}$. We use the Landau-Lifshitz sign conventions with a ``mostly minus'' metric.

This is a simple generalization of the action for self-gravitating non-topological solitons like boson stars or Q-stars \cite{FriedbergLeeSirlin1976,Coleman1985,Lynn1988,Jetzer1992,Lee-Pang1991,SchunckMielke2003} which are obtained from this action for $\beta=0$ with a potential function which may contain only a mass term, or be a higher order polynomial like the ordinary $|\Phi|^4$ potential or a non-monotonic function as in the case of Q-balls:
\be \label{potential}
U(|\Phi|) = \frac{m^2}{2} |\Phi|^2 +  \frac{\lambda}{4} |\Phi|^4\;\;\;  , \;\;\;\;\;
U_Q(|\Phi|) = \frac{m^2}{2} |\Phi|^2 - \frac{\zeta}{4} |\Phi|^4 + \frac{\nu}{6m^2} |\Phi|^6
\ee
In this work we will take the simplest choice of a mass term only.

The non-minimal coupling term to the Einstein tensor is known to introduce in the field equations of the pure scalar case derivatives of order not higher than two \cite{Horndeski1974}. It is easy to show (as we will do shortly) that the gauged version still has this property.

The field equations derived from this action are (for the matter fields): \begin{eqnarray}\label{GeneralFieldEqsScalar}
D_{\mu} D^{\mu} \Phi - 2\beta G^{\mu \nu} D_{\mu} D_{\nu} \Phi+ \frac{\Phi}{|\Phi|} \frac{dU}{d|\Phi|}=0\\\label{GeneralFieldEqsVector}
\nabla_\mu F^{\mu\nu} = J^\nu -  2\beta G^{\mu \nu} J_{\mu}  = {\cal J}^\nu
\end{eqnarray}
where
\be \label{Current}
J_{\nu} = -\frac{ie}{2}\left(\Phi^{*} D_{\nu} \Phi - \Phi D_{\nu} \Phi^{*}\right)
\ee
is the ordinary current of the minimally-coupled case. Note that  $J^{\nu}$ is still conserved although the full source of the electromagnetic field is ${\cal J}^\nu$ which is of course conserved as well.

The gravitational (``Einstein'') equations can be written in various forms. Probably the most compact is:
\begin{eqnarray}\label{GeneralFieldEqsEinst}
G_{\mu \nu} + \kappa T^{(matter)}_{\mu \nu} -\beta\kappa \left( X_{\mu\nu} +
Yg_{\mu\nu} + Z_{\mu \nu}\right)=0
\end{eqnarray}
where \footnote{By writing terms like $D_\mu \Phi^{*} D_{\nu}\Phi $ we mean
$(D_\mu \Phi^{*})(D_{\nu}\Phi)$ and not  $D_\mu (\Phi^{*} D_{\nu}\Phi )$.}
\begin{eqnarray}\label{Tmunu}
T^{(matter)}_{\mu \nu} = \frac{1}{2}\left(D_\mu \Phi^{*} D_{\nu}\Phi +  D_{\mu}\Phi D_{\nu}\Phi^{*}\right)
-g_{\mu\nu} \left( \frac{1}{2} D_{\rho}\Phi^* D^{\rho}\Phi - U(|\Phi |)
-\frac{1}{4} F_{\rho\sigma}F^{\rho\sigma} \right) + F_\mu^\rho F_{\rho\nu}
\end{eqnarray}
\begin{eqnarray}\label{Xmunu}
X_{\mu \nu} = g_{\mu\nu}G^{\rho\sigma} D_{\rho}\Phi^*D_{\sigma}\Phi +
\frac{R}{2}\left(D_\mu \Phi^{*} D_{\nu}\Phi + D_{\mu}\Phi  D_{\nu}\Phi^{*}\right) +
R_{\mu \nu} D_{\rho}\Phi^* D^{\rho}\Phi- \;\;\; \\ \nonumber
R_{\rho\mu}\left(D^\rho \Phi^{*} D_{\nu}\Phi +  D^\rho\Phi D_{\nu}\Phi^{*}\right) -
R_{\rho\nu}\left(D^\rho \Phi^{*} D_{\mu}\Phi +  D^\rho\Phi D_{\mu}\Phi^{*}\right)
\end{eqnarray}
\begin{eqnarray}\label{Y}
Y =\frac{1}{2}\nabla_\rho \nabla_\sigma\left(D^\rho \Phi^{*} D^{\sigma}\Phi +
D^{\rho}\Phi  D^{\sigma}\Phi^{*}\right)- \nabla_\rho \nabla^\rho\left(D_\sigma \Phi^{*} D^{\sigma}\Phi\right)
\end{eqnarray}
\begin{eqnarray}\label{Zmunu}
Z_{\mu \nu} =\frac{1}{2} \nabla_\rho \nabla^\rho \left(D_\mu \Phi^{*} D_{\nu}\Phi + D_{\mu}\Phi  D_{\nu}\Phi^{*}\right)+ \nabla_\mu \nabla_\nu \left(D_{\rho}\Phi^* D^{\rho}\Phi \right)-
\hspace{3cm} \\ \nonumber
\frac{1}{2} \nabla^\rho \left[ \nabla_\mu \left(D_\rho \Phi^{*} D_{\nu}\Phi +  D_\rho\Phi D_{\nu}\Phi^{*}\right) +  \nabla_\nu \left(D_\rho \Phi^{*} D_{\mu}\Phi +  D_\rho\Phi D_{\mu}\Phi^{*}\right)\right]
\end{eqnarray}

We notice that only the last two terms, $Y$ and $Z_{\mu \nu}$ contain higher order derivatives and it is easy to get convinced that by expanding the derivatives and using the commutation relations of the covariant derivatives, all the third order derivatives of the scalar field drop out completely and only second derivatives (and lower) are left. The final result is quite lengthy and we do not present it here because  the form of Eqs. (\ref{GeneralFieldEqsEinst})-(\ref{Zmunu}) is more convenient even for concrete calculations.

\section{Field Equations - Spherical Symmetry}
\label{Spherical}
\setcounter{equation}{0}

We assume a static spherically symmetric  form of metric given by the line element
\be \label{spher-metric}
ds^2=g_{\mu\nu}dx^\mu dx^\nu= A(r)^2 dt^2-  B(r)^2 dr^2- L(r)^2 d \Omega_2^2
\ee
while for the scalar and vector fields we assume the radial ``electric'' configuration
\be \label{ElectricAnsatz}
\Phi=f(r)e^{i\omega t} \;\;\;  , \;\;\;\;\; A_\mu dx^\mu = a_0 (r) dt
.\ee

By choosing the gauge $L(r)=r$ we obtain 4 independent field equations for the 4 functions
$f(r)$, $a_0 (r)$, $A(r)$ and $B(r)$. The first two for the scalar and vector fields:
\begin{eqnarray}\label{Eqf}
 \left[1+\frac{2 \beta }{r^2} \left(1-\frac{1}{B^2}-
 \frac{2 r A'}{B^2 A}\right)\right] f'' \hspace{8cm}\\ \nonumber
   + \left[\frac{2}{r}+\frac{A'}{A}-
   \frac{B'}{B}+\frac{2 \beta}{r^2}
   \left(\frac{A'}{A}-\frac{B'}{B}-\frac{1}{B^2}\left(\frac{2 r A''}{A}-\frac{6 r A' B'}{A B}+
   \frac{3A'}{A}-\frac{3B'}{B}\right)\right)\right]  f'\\  \nonumber
   +\frac{B^2 }{A^2} \left[1+\frac{2 \beta  }{r^2}\left(1-\frac{1}{B^2}+\frac{2 r
   B'}{B^3}\right) \right] (\omega -e a_0)^2 f-B^2 U'(f) = 0
\end{eqnarray}
\begin{eqnarray}\label{Eqa0}
 a_0 ''+ \left(\frac{2}{r}-\frac{A'}{A}-\frac{B'}{B}\right)a_0 ' +
 B^2 \left[1+\frac{2 \beta}{r^2} \left(1-\frac{1}{B^2}+\frac{2 r B'}{B^3}\right)\right]
 e f^2 (\omega -e a_0) = 0
\end{eqnarray}
Note the repeating appearance in the equations of the components of the Einstein tensor:
\begin{eqnarray}\label{EinsteinTensor}
G_0^0=-\frac{1}{r^2} \left(1-\frac{1}{B^2}+\frac{2 r B'}{B^3}\right) \; , \hspace{0.25cm}
 G_r^r=-\frac{1}{r^2} \left(1-\frac{1}{B^2}-\frac{2 r A'}{ B^2 A}\right) \; , \hspace{1cm}
 \\ \nonumber
 G_\theta^\theta=  G_\phi^\phi= \frac{1}{B^2} \left(\frac{A''}{A}+\frac{A'}{rA}-\frac{B'}{rB}-\frac{A'}{A}\frac{B'}{B}\right)
\end{eqnarray}
The other two equations are:
\begin{eqnarray}\label{EqA}
\left(1-\frac{\beta \kappa  \left(\omega -ea_0
   \right)^2 f^2}{A^2}\right)\left(1-\frac{1}{B^2}+\frac{2 r B'}{B^3}\right)  \hspace{9cm}\\ \nonumber
   -\kappa  r^2 \left[\frac{\left(a_0'\right)^2}{2 A^2 B^2}+\frac{ \left(\omega
   -ea_0 \right)^2 f^2}{2 A^2}+\left(1+\frac{2 \beta  }{r^2}\left(1+\frac{1}{B^2}-\frac{6 r
   B'}{B^3}\right)\right)\frac{\left(f'\right)^2 }{2 B^2}+U(f)\right]-\frac{4 \beta  \kappa r f' f''}{B^4}  = 0
\end{eqnarray}
\begin{eqnarray}\label{EqB}
\left(1-\frac{\beta \kappa  \left(\omega -ea_0
   \right)^2 f^2}{A^2}\right)\left(1-\frac{1}{B^2}-\frac{2 r A'}{B^2 A}\right)  \hspace{9cm}\\ \nonumber
   +\kappa  r^2 \left[-\frac{\left(a_0'\right)^2}{2 A^2
   B^2}+\left(1+\frac{4 \beta }{r^2} \left(1-\frac{1}{B^2}\right)\right)\frac{ \left(\omega -ea_0 \right)^2 f^2}{2 A^2}+\left(1+\frac{2 \beta }{r^2} \left(1-\frac{3}{B^2}-\frac{6 r A'}{B^2 A }\right)\right)\frac{\left(f'\right)^2 }{2 B^2}-U(f)\right] \\ \nonumber
   -\frac{2 \beta  \kappa  r \left(\left(\omega -ea_0\right)^2 f^2 \right)'}{A^2 B^2} = 0
\end{eqnarray}

There is a fifth equation, the $G_\theta^\theta$ or $G_\phi^\phi$ equation which is not independent. We do not write it here since it is quite cumbersome, but we use it as a consistency check of our solutions.

These equations can be derived from the following effective Lagrangian:
\begin{eqnarray} \label{Leff}
L_{eff}= \frac{1}{\kappa }\left(\frac{A \left(L'\right)^2}{B}+\frac{2 L L' A'}{B}+A B\right)+ A B L^2 \left[- \left(1+2 \beta
   \left(\frac{1}{L^2}-\frac{2 L' A'}{B^2 L A}-\frac{(L')^2}{B^2 L^2}\right)\right)\frac{\left(f'\right)^2 }{2 B^2} \nonumber  \right. \\ \left.
   +\left(1+2 \beta  \left(\frac{1}{L^2}-\frac{2 L' A'}{B^2 LA}+\frac{(L')^2}{B^2 L^2}\right)\right)\frac{f^2 \left(\omega -ea_0\right)^2 }{2 A^2} +\frac{\left(a_0'\right)^2}{2 A^2 B^2}-U(f)\right] \nonumber \\
   +\frac{4\beta  L L'  }{A B} \left(f f' \left(\omega
   -ea_0\right)^2-e f^2  \left(\omega -ea_0\right)a_0'\right)
     \end{eqnarray}
and using the gauge condition $L(r)=r$ at the last step.

It is worth noting already at this point, the factor  which multiplies the Einstein tensor components in both Einstein equations. It seems to give rise to an effective field-dependent gravitational constant of
$\kappa/\left(1-\beta \kappa  \left(\omega -ea_0 \right)^2 f^2/A^2\right)$ which strengthens gravity with increasing $\beta$.

\section{Physical parameters}
\label{PhysPar}
\setcounter{equation}{0}

We will characterize the solutions by their total mass, electric charge and particle number. The ADM mass may be
read of the asymptotic decay of the metric $1/B^{2}(r) = 1 - 2 {\cal G} M /r + o(1/r)$. We will define also the mass function ${\cal M}(r)$ by
\be \label{MassFctn}
\frac{1}{B^{2}(r)} = 1 - \frac {2  {\cal M}(r)}{r}
\ee
So we can also write ${\cal G} M = {\cal M}(\infty)$. Using the mass function simplifies some of the equations; e.g. $G_0^0=-2 {\cal M}' /r^2$.
As usual, we can also calculate the total mass by integration of the energy density $T_0^0$:
\be \label{InertialMass}
M =  {\cal M}(\infty)/{\cal G} = 4 \pi \int_0^\infty dr r^2  T_0^0
\ee
But the non-minimal coupling makes it difficult to disentangle the matter terms from the curvature, so the
practical use of this expression is limited. The simplest will be to calculate the mass from the asymptotic form of the metric.

A second important characteristic of the solutions is the local $U(1)$ charge.
It is readily obtained from the time component of the conserved total current defined in Eq. (\ref{GeneralFieldEqsVector}) that in the spherically-symmetric case reads:
\be \label{SphCurrent0}
{\cal J}_0 =  e \left(\omega -ea_0\right) f^2 (1-2\beta G_0^0 )=
 e  (\omega -e a_0) f^2 \left(1+\frac{4 \beta {\cal M}'}{r^2} \right)
\ee
The total electric charge is therefore:
\be \label{TotalCharge}
Q=  4 \pi e \int_0^\infty dr  \frac{r^2 B}{A} (\omega -e a_0) f^2 \left(1+\frac{4 \beta {\cal M}'}{r^2} \right)
\ee
 We assume of course that the integral converges for the localized solution we are after. Without loss of generality we will take $\omega>0$, so $Q>0$ as well.

 These two quantities, mass and charge determine the asymptotic behavior of the localized solution that we seek, i.e. the exterior Reissner-Nordstr\"{o}m solutions:
\begin{eqnarray} \label{RN}
ds^2= \left(1-\frac{\kappa M}{4\pi r}+\frac{\kappa Q^2}{2(4\pi)^2 r^2}\right) dt^2-  \left(1-\frac{\kappa M}{4\pi r}+\frac{\kappa Q^2}{2(4\pi)^2 r^2}\right)^{-1} dr^2 - r^2 d \Omega_2^2 \;\;\;\;\; ; \;\;\;\;\;
a_0 = \frac{Q}{4\pi r}
\end{eqnarray}
So, we may extract also the charge from the asymptotic behavior of the solutions.

The total electric charge is carried in our case by $N$ particles of mass $m$ and charge $e$. We can therefore define a particle density and particle number by
\be \label{ParticleNumber}
{\cal \rho}_N = (\omega -e a_0) f^2 \left(1+\frac{4 \beta {\cal M}'}{r^2} \right)
\;\;\ ; \;\;\;
N =  4 \pi  \int_0^\infty dr  \frac{r^2 B}{A} (\omega -e a_0) f^2 \left(1+\frac{4 \beta {\cal M}'}{r^2} \right) .
\ee
Thus, we can use the mass to particle number ratio, or more precisely, $M/mN$  in order to study the stability of these solutions against ``fission'' into a number of smaller stable structures: this ratio $M/mN$  needs to be less than 1 in order for the solutions to be stable against this kind of processes.
The condition $M/mN<1$ is however not sufficient to guarantee the stability
under linear perturbations \cite{Jetzer1992}. We will postpone this issue to a future work.
Note that unlike ${\cal J}_0$ and $Q$, the corresponding particle density and number do not vanish for the pure scalar system, i.e. when $e=0$ and $A_\mu=0$, and may be still used in order to characterize the solutions.

\section{Rescaling and dimensionless variables}
\label{Rescaling}
\setcounter{equation}{0}

The field equations depend a priori on the 3 parameters $m$, $e$, $\beta$ and Newton's constant $\cal {G}$. The equations can be written in a dimensionless form which reveals a dependence on two independent parameters only: we replace $r$ by $x=mr$ and scale the fields $f$ and $a_0$ by a factor $\mu=1/\sqrt {\kappa}$ . Then the equations depend on the two dimensionless parameters
\be
 \label{RescPar}
     \overline \beta = \beta m^2 \ \ , \ \ \overline e =  \frac{e}{\sqrt{\kappa m^2}} \; .
\ee

Note that the above rescaling further implies that the frequency is
 rescaled according to $\overline \omega = \omega/m$.

The dimensionless version of the field equations (\ref{Eqf})-(\ref{EqB}) are obtained formally by taking $m=1$ and $\kappa=1$ and replacing $r$ by the dimensionless radial variable $x$.

We also comment that the rescaled version of the charge, particle number and mass are obtained by the same way exactly. They are related to one another by $\bar{M}=M m/\mu^2=\kappa  m M$ , $\bar{Q}=Qm^2/\mu^2=Q\kappa m^2$ and $\bar{N}=Nm^2/\mu^2=N\kappa m^2$. The stability ratio may be calculated both ways: $M/mN = \bar{M}/\bar{N}$.

The mass function ${\cal M}(r)$ which we defined above has a dimension of length and is rescaled therefore similarly as $ \bar{{\cal M}} = m{\cal M}$. The total mass may therefore be expressed in terms of
$\bar{{\cal M}}(\infty)$ or  $\bar{M}$ as $M=8\pi \bar{{\cal M}}(\infty) m_{Pl}^2/m = \bar{M} m_{Pl}^2/m$. Similarly the rescaled particle number $\bar{N}$  may be used to express the total particle number as $N=8\pi \bar{N} (m_{Pl} /m)^2$.

In the following, we will present the results in terms of the  dimensionless variables and omit the 'bar' of $\overline \omega$,  $\overline e$,  $\bar {M}$ and so on and use $r$ to represent $x=mr$.

\begin{figure}[t!]
\begin{center}
{\includegraphics[width=8cm]{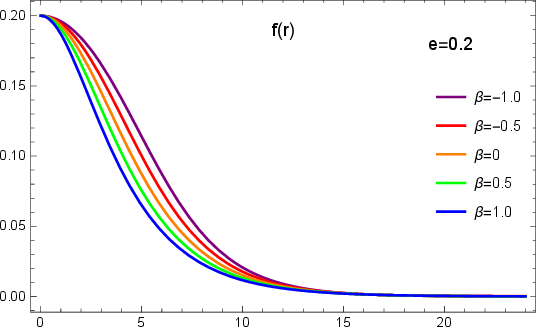}}
{\includegraphics[width=8cm]{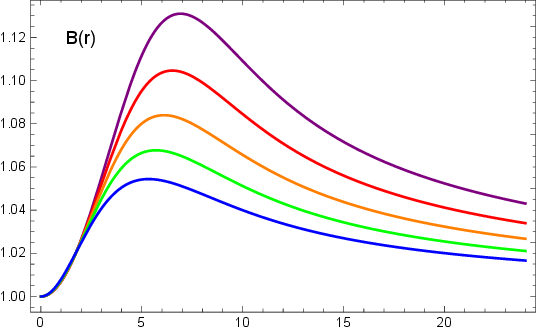}}
{\includegraphics[width=8cm]{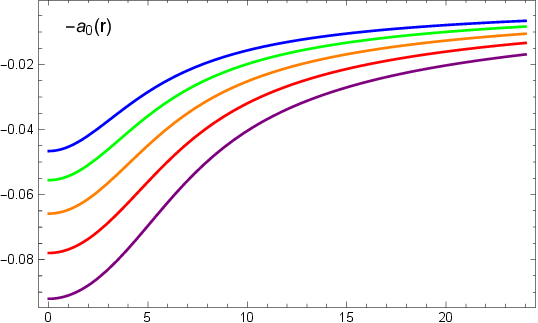}}
{\includegraphics[width=8cm]{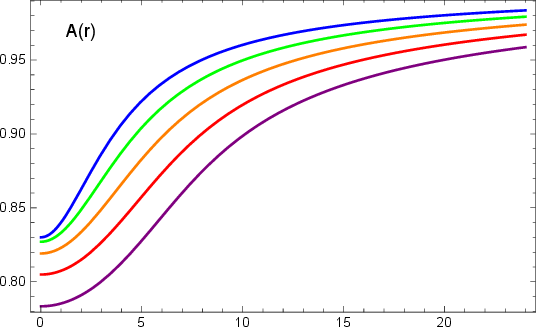}}
\caption{Solution profiles for $e=0.2$ and $\beta=0, \pm 0.5,  \pm 1$. The colors vary from purple which corresponds to $\beta= -1.0$ to blue which corresponds to $\beta= 1.0$. For the values of mass and particle number, see text. The way in which they increase is obvious from the depth of the gravitational ``potential well'' represented by $A(r)$.\label{FourProf}}
\end{center}
\end{figure}

\section{Solutions and their properties}
\setcounter{equation}{0}

Although in some special cases of this theory (e.g. the neutral system with shift symmetry) solutions were obtained explicitly   (``analytically'' - see e.g. \cite{BabichevEtAl2016}), we found it unlikely to achieve this in the present system we study here. We therefore revert to numerical analysis in order to  construct our solutions. We used in particular the routine COLSYS. This solver
developed by Ascher, Christiansen and Russell \cite{colsys} uses a collocation method
for boundary-value ordinary differential equations.
The set of non-linear coupled differential equations (\ref{Eqf})-(\ref{Eqa0}), (\ref{EqA})-(\ref{EqB}) was solved using the damped Newton method of quasi-linearization. At each iteration step a linearized problem
is solved by using a spline collocation at Gaussian points. This method works efficiently
when the initial approximate solution is close to the true solution. So we could use
known profiles of uncharged, minimally-coupled solutions and proceed by increasing gradually
the various parameters entering in the equations.
More details on the method are given, e.g. in the Appendix of \cite{Brihaye:1999kt}.
In this paper, the used mesh includes typically $10^3$ points and the relative accuracy
of the solutions is typically of the order of $10^{-8}$.

Within our ansatz (\ref{spher-metric})-(\ref{ElectricAnsatz}) and the gauge choice $L(r) = r$,
the unknown functions $B,A,f,a_0$ appear with maximal derivatives
$B', A'', f'', a_0''$  in the  system of reduced equations (\ref{Eqf})-(\ref{Eqa0}), (\ref{EqA})-(\ref{EqB}).
These maximal derivatives appear linearly, so that the equations are
symbolically
\be
      M (B',A'',f'',a_0'')^T = K
\ee
 where $M$ represents a {\it non-diagonal} $4\times 4$ matrix and
 $K$ a  $4\times 1$ matrix (vector)  whose entries depend on the lower
 derivatives.
 The quantity $\Delta(r) \equiv {\rm det} M$ plays a major role in
 the existence of regular solutions over space-time.
 Indeed zeroes of $\Delta(r)$  correspond to singular points of the system.
 In particular, the occurrence
 of such a point renders impossible the construction of numerical
 solutions on $r \in [0,\infty]$. This plays a major role in the
 understanding of the pattern of the solutions.
 For completeness, we write here the expression of $\Delta$ in terms of the fields  (we write for short $w=\omega -ea_0 $). In terms of the paramterization used here, $\Delta (r)$ reads:
\begin{eqnarray} \label{Delta}\nonumber
\Delta (r)=  \left[\left(1+\frac{2 \beta }{r^2}\right)B^2 -\frac{2 \beta }{r^2}\right] \left(1-\beta  \kappa\frac{  w^2 f^2 }{A^2}\right)^2+2 \beta  \kappa\left[ \left(1+\frac{2 \beta }{r^2}\right)B^2 + \frac{4 \beta  A'}{r A}\right] \left(1-\beta  \kappa\frac{ w^2 f^2}{A^2}\right)\frac{  f'^2 }{B^2}+\\ \nonumber
 \frac{4 \beta ^2 \kappa }{r^2 }\left(1-\beta  \kappa \left(5-16 \beta  e^2 f^2\right)\frac{ w^2 f^2 }{A^2}\right)\frac{  f'^2 }{B^2}+ \frac{6 \beta ^3 \kappa ^2 }{r^2 }\left(1+\frac{2 r A'}{A}\right)\frac{f'^4}{ B^4}+\beta ^2 \kappa ^2 \left(1+\frac{2 \beta }{r^2}\right)\frac{ f'^4}{B^2}+\\
 \frac{16 \beta ^3 \kappa ^2
   }{r}\frac{ w^2
   f f'^3}{ A^2 B^2}+ \frac{4 \beta ^2 \kappa }{r }\frac{ w^2 }{ A^2} \left[\left(2-\beta \kappa\frac{ w^2 f^2}{A^2}\right)\frac{f^2 A' }{A}+4  \left(1-\beta  \kappa\frac{ w^2 f^2}{A^2}\right)f f'\right]-\frac{4 \beta  A'}{r A}
\end{eqnarray}

\begin{figure} [t!]
\begin{center}
{\includegraphics[width=10cm]{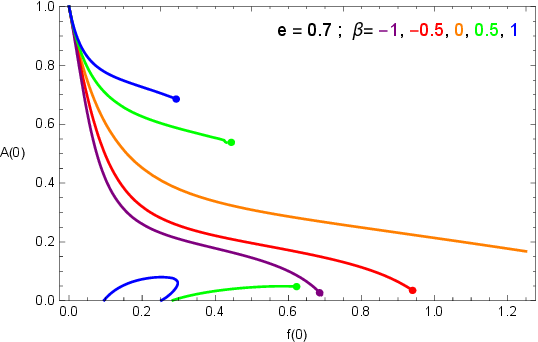}}
\caption{Families of solutions for $e=0.7$ and 5 values of $\beta$: $\beta= \pm 1, \pm 0.5$ and for comparison the minimally coupled case $\beta = 0$. Note the two branch structure for  $\beta>0$ and the different behavior of branch 2 families appearing at the lower part of the figure.
The ``bullets'' stress the end point solutions without back-bending, i.e. $\Delta = 0$ solutions. The other end points at $A(0)=0$ are extremal Reissner-Nordstr\"{o}m solutions - see text.
\label{Central_f_vsAe07}}
\end{center}
\end{figure}

In Fig. \ref{FourProf} we present some profiles of the solutions for\footnote{recall: these are the dimensionless versions of the parameters.}  $\beta= -1.0 , -0.5 , 0 , 0.5 , 1.0  $ - all with the same $ e = 0.2$ and the same central value of the scalar field, $f(0)=0.2$. The masses are in this order: $M=$ 24.428, 19.514, 15.521, 12.339, 9.801; Particle numbers: $N=$ 25.472, 20.141, 15.879, 12.525, 9.879. Note that the mass and particle number decrease with $\beta$ which is in accord with the fact that gravity becomes stronger with increasing $\beta$, so a smaller number of particles can form bound states.

The dependence on the coupling constant $e$ is expected to cause the opposite effect since increasing its value amounts to a stronger electrostatic repulsion. This is indeed the general behavior as we will see in what follows. The electrostatic repulsion is also responsible for a critical value of $e$ which in terms of the dimensionless parameter reads $\overline e_{cr} = 1/\sqrt{2}$. Actually, this bound was obtained \cite{Jetzer1992} for minimal coupling ($\beta=0$), but it still holds here with a weak dependence on $\beta$. We thus suppose that the Jetzer-Liljenberg-Skagerstam  bound \cite{JetzerEtAl1993} is valid also for $\beta\neq 0$ if we interpret the gauge fields as those of ordinary electromagnetism.

For a given set of $\beta$ and $e$, the solutions comprise a one-parameter set parametrized by $f(0)$ or alternatively by $\omega$  or $A(0)$, although in some regions of parameter space the correspondence is not one to one as will become evident from the plots below.

In other areas of parameter space we find a second branch of solutions with significantly different behavior with respect to the first (``ordinary'') branch. Fig. \ref{Central_f_vsAe07} shows the general structure in the $(f(0),A(0))$ plane for $e=0.7$ and five values of $\beta$: $\beta= \pm 1, \pm 0.5$ and, for better orientation, the minimally coupled case $\beta = 0$. The two-branch structure appears for $\beta>0$ only and above a certain value of $e$.

\begin{figure}[t!]
\begin{center}
{(a)\includegraphics[width=8cm]{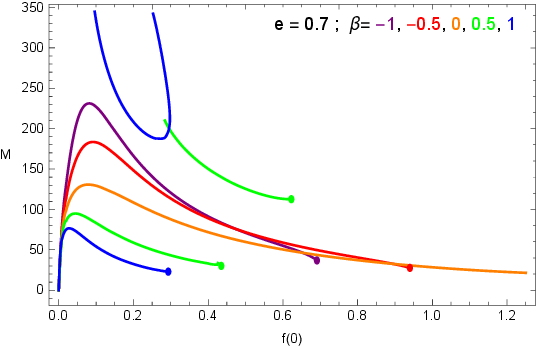}}\;\;\;\;\;\;\;\;\;\;\;\;(b)
{\includegraphics[width=6cm]{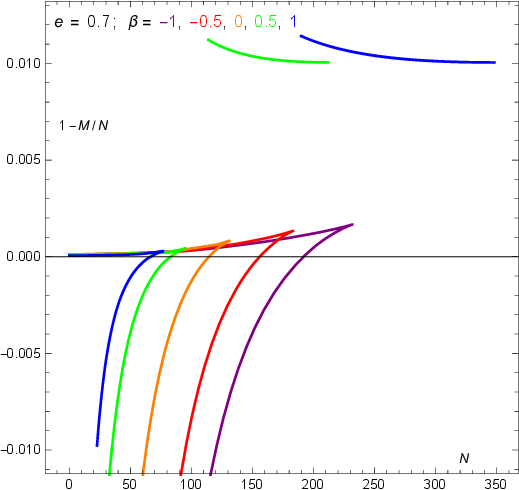}}
\caption{Mass and binding energy per particle for $e=0.7$ and 5 values of $\beta$: $\beta= \pm 1, \pm 0.5$ and  $\beta = 0$. (a) Mass versus the central value of the scalar field; (b) Binding energy per particle as a function of particle number. The branch 2 curves are the upper ones at both figures. The ``un-bulleted'' ends of the branch 2 curves in (a) represent the extremal Reissner-Nordstr\"{o}m solutions. These correspond to the $A(0)=0$ points of  Fig. \ref{Central_f_vsAe07}.
\label{PropertiesVarBet+e07}}
\end{center}
\end{figure}

In order to understand the situation better, we calculate the main properties of the solutions: mass, particle number and their ratio which is in 1 to 1 correspondence with the binding energy per particle $(N-M)/N=1-M/N$ . We study their dependence on the parameters of the theory $\beta$ and $e$ as well as their behavior within a given family corresponding to specific $(\beta, e)$ and parametrized by $f(0)$. Fig. \ref{PropertiesVarBet+e07} contains further aspects of the two-branch structure of the theory:
the mass as a function of the central scalar field $f(0)$ and the dependence of binding energy per particle on the particle number.

Branch 1 of the solutions is quite similar to that of charged ordinary (i.e. $\beta=0$) boson stars for both signs of $\beta$: it starts at $f(0)=0$ with $M=N=0$, then the mass and particle number grow steeply, attain (together) a maximum and then decrease. Table \ref{MaxMass} summarizes the characteristics of the maximal mass configurations with $e=0.7$.

\begin{table}[b!]
  \centering
\begin{tabular}{@{}ccccccc@{}}
\hline \hline
$\beta$   & -1 & -0.5 & 0 & 0.5 & 1 \\
\hline
$ M_{max }$   & 231.23 & 183.54 & 130.77 & 94.944 & 76.64 \\
\hline
$ N_{max }$   & 231.62 & 183.79 & 130.88 & 94.986 & 76.661 \\
\hline
$1-M_{max}/N_{max}$   & $1.6674\times 10^{-3}$ & $1.3533\times 10^{-3}$ & $0.81708\times 10^{-3}$ & $0.43723\times
   10^{-3}$ & $0.28052\times 10^{-3}$ \\
\hline
$f(0)$    & 0.08183 & 0.09153 & 0.07858 & 0.04484 & 0.02762 \\
\hline
$A(0)$    & 0.47475 & 0.51999 & 0.65578 & 0.80729 & 0.87837 \\
\hline \hline
\end{tabular}
  \caption{The characteristics of the maximal mass configurations with $e=0.7$ and various values of $\beta$.}\label{MaxMass}
\end{table}

However, unlike the case of ordinary charged boson stars, the solutions hit an end point beyond which no solutions exist. This end point  has a similar origin as reported for the ungauged counterpart   \cite{BrihayeEtAl2016} where the function $\Delta (r)$ (see Eq. (\ref{Delta})) was found to develop a zero at the end point excluding solutions beyond that point.

\begin{figure} [t!]
\begin{center}
{\includegraphics[width=9cm]{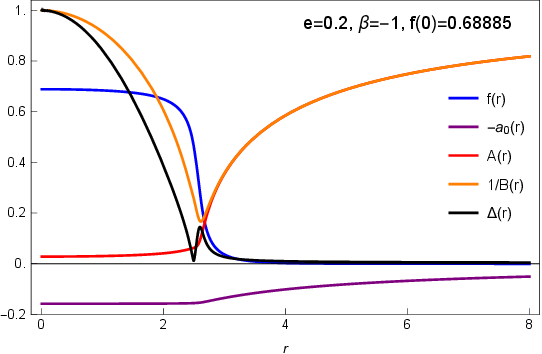}}
\caption{Profiles for the end point solution for $e=0.2$ and $\beta=-1$ together with $\Delta(r)$ whose zero is responsible for this being an end point solution.
\label{Plots+Delta}}
\end{center}
\end{figure}

Fig. \ref{Plots+Delta} contains the profiles of a typical end point solution of branch 1 together with $\Delta (r)$ and its visible zero.

Branch 2 solutions exist only in restricted regions of parameter space: for a given $e$, $\beta$ should be larger than a certain value. For given $e$ and $\beta$, branch 2 starts at a minimal non-zero $f(0)$ and $A(0)=0$ and has also a maximal $f(0)$. This maximal $f(0)$ point can be either:\\
1) An end point of a similar kind of branch 1 where solutions cease to exist because $\Delta (r)$ gets a zero. In this case $A(0)>0$ at that point.\\
2) A turning point in which back-bending occurs and solutions still exist down to some final $f(0)$ which is an end point because $A(0)=0$.\\

To sum up, there are two kinds of end points of curves in parameter space as is seen clearly in Fig. \ref{Central_f_vsAe07} and \ref{PropertiesVarBet+e07}: The first kind are the ``bulleted'' which correspond to the $\Delta=0$ termination condition. These appear on the $\beta\neq 0$ branch 1 curves and the higher $f(0)$ end point of the $\beta=-0.5$ branch 2 curve. End points of the second kind are the ``un-bulleted'' two end points of the $\beta=-1$ branch 2 curve and the lower $f(0)$ end point of the $\beta=-0.5$ branch 2. These correspond to the $A(0)=0$ termination condition.

\begin{figure}[b!]
\begin{center}
{\includegraphics[width=11cm]{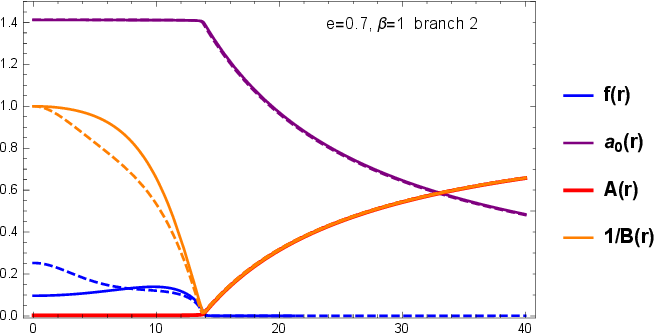}}
\caption{The two end-point (``almost BH'') solutions for $e=0.7$ and $\beta= 1.0$. Both solutions have $N=348.2$ and $M=344.7$ and the same external structure (with $A(r)$ and $1/B(r)$ almost identical), but very different internal structure: $f(0)=0.096255$ and $f''(0)>0$ - solid; $f(0)=0.2525$ and $f''(0)<0$ - dashed.
\label{BHProfs}}
\end{center}
\end{figure}

At the end points of branch 2 where $A(0)\rightarrow 0$ and the scalar field becomes confined within a ball with a sharp boundary at $r=r_S$, we find that the solutions approach extremal Reissner-Nordstr\"{o}m black hole solutions (ERN or ERNS in the following) which have identical mass and particle number, but correspond to very different internal structure. In our parametrization the ERNS have the characteristic relation $M=\sqrt{2} Q$ or in terms of particle number, $M=\sqrt{2} e N$. Fig. \ref{BHProfs} shows for $e=0.7$ and $\beta= 1.0$ the identical exterior solutions of the end-point solutions that correspond to the two different internal structures. The $M$ and $N$ values of these solutions $M=344.7$ and  $N=348.2$ satisfy this relation very nicely. Another characteristic of the ERNS, is the equality $A(r)=\omega -ea_0 (r)$ which we do not show in our plots, but was checked to hold very well.

\begin{figure}[t!]
\begin{center}
{\includegraphics[width=8cm]{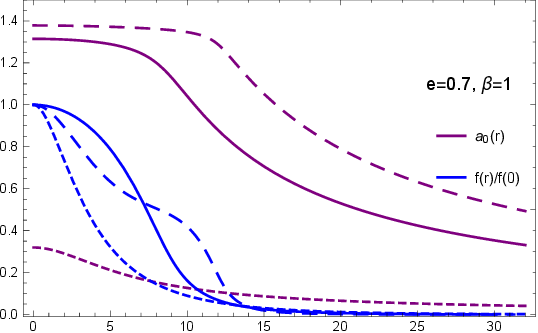}}
{\includegraphics[width=8cm]{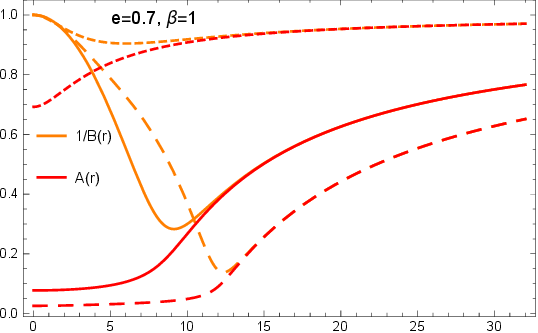}}
\caption{Profiles of three solutions with the same $f(0)$ for $e=0.7$ and $\beta= 1.0$. All solutions have $f(0)=0.28$, but very different behavior: branch 1 - densely dashed; branch 2 higher mass - dashed; branch 2 lower mass - solid. The corresponding masses are: 23.685, 280.484, 188.084. \label{MoreProf}}
\end{center}
\end{figure}

We like to point out that the internal structure of this ``almost ERN'' boson star is not that of a charged conducting ball, as might be inferred from the fact that its gauge potential ($a_0$) is practically purely Coulombic outside the sphere $r=r_S$ and constant inside. The reason is that the metric components play inside the ball the role of dielectric functions thus preventing all the charges from accumulating on the surface and keeping a non-zero internal electric field.

\begin{figure}[b!]
\begin{center}
{\includegraphics[width=12cm]{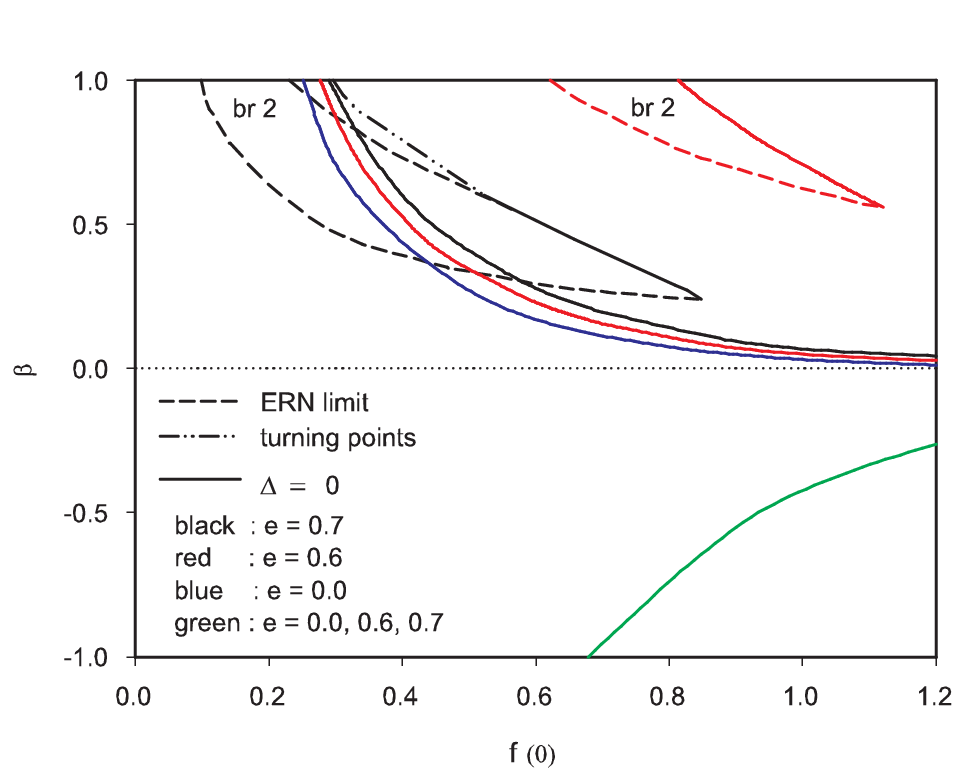}}
\caption{The branch structure of parameter space for $e=$ 0, 0.6, 0.7  and $-1\leq \beta \leq 1$. Note that branch 1 extends from the vertical axis at $f(0)=0$ up to the curves. Branch 2 regions are indicated explicitly. For further explanation - see text.
\label{DomainStructure}}
\end{center}
\end{figure}

Finally, we note that the property of sharp boundary which confines the scalar field in its interior is shared also by the charged compact boson stars found by Kunz {\it et al.} - see e.g. \cite{KumarEtAl2017}.

Another aspect of the rich structure of parameter space of this system is demonstrated in Fig. \ref{MoreProf} where we present three profiles with the same value of the central scalar field $f(0)=0.28$. The difference in shape is very noticeable as is reflected by the three corresponding mass values:  23.685, 188.084, 280.484.

As a summary of the branch structure of this theory we present Fig. \ref{DomainStructure} which illustrates the structure of parameter space in the $(f(0),\beta)$ plane for $e=$ 0.6, 0.7 and for comparison with previous work also $e=0$. The $e=0$ case was studied in Ref. \cite{BrihayeEtAl2016}. Note that the line for the $\beta<0$ solutions is the line of end points for any value of $e$: the maximal $f(0)$ depends on $\beta$ but not on $e$.

From this plot, together with the other ones, we collect the following properties:\\
1) Branch 1 solutions are the more common. They are stable for the smaller $f(0)$ region up to some maximal value which decreases with $\beta$.\\
2) The curve in the $(f(0),\beta)$ plane which defines the boundary of the branch 1 region is fixed by the end point solutions with a zero of $\Delta (r)$.\\
3) Branch 2 solutions exist in limited region of parameter space, but can extend to values of $f(0)$  higher than in branch 1. \\
4) The left side boundary of branch 2 is defined by the condition $A(0)=0$ which corresponds to the ERNS.\\
5) The right-hand-side of the boundary may consist either of purely $\Delta =0 $ solutions as for $e=0.6$, or of a mixture where the lower $\beta$ have similar origin, but the higher $\beta$ part is a line of turning points beyond which solutions continue until $A(0)=0$ is reached at the ERNS.\\
6) The branch 2 solutions are much more massive than those of branch 1. They are also
all stable and much more so than the stable branch 1 solutions.\\
7) For both branches the effect of increasing $\beta$ (stronger gravity) is to decrease the mass and particle number.
On the other hand increasing $e$ increases also the mass and particle number in order to overcome stronger repulsion.



\section{Conclusion}
\setcounter{equation}{0}

Charged boson stars \cite{Jetzer1992,SchunckMielke2003} are known for quite a long period.
The initial goal of this paper was to investigate the deformation of these solutions for a generalization of the kinetic part of the underlying scalar field.
More precisely by means of the so called ``John'' sector of
the Horndeski theory of scalar-tensor gravity.
The additional (non-minimal) coupling is determined by an additional
coupling constant which we denote $\beta$.

Our numerical analysis of the equations has indeed confirmed that the charged boson stars, characterized by the three parameters
$e$, $\beta$ and $f(0)$ exist for values of the coupling constant $e$ not higher than the critical value of the minimal coupling $e_{cr}=\sqrt{\kappa m^2 /2}$ and in a finite interval of the non minimal coupling parameter
$\beta \in [\beta_m, \beta_M]$ where $\beta_m, \beta_M$ depend on $e$ and of $f(0)$.

However, a deeper investigation of the solutions has revealed the existence of an additional family of solutions
referred to as 'branch 2-solutions' while the above-mentioned solutions belong to 'branch 1'.

Likely, these branch 2-solutions exist due to the high degree of non-linearity of the classical equations.
For instance, they exist for sufficiently large
values of both  $\beta$ and $e$. Interestingly they posses a few remarkable properties; for instance~:
(i) their binding energy is larger than any of the solutions of the branch 1;
(ii) on a part of their domain of existence they terminate into extremal Reissner-Nordstr\"{o}m black holes.
The existence of generic black holes with scalar hair seems, however, incompatible with the equations.

The understanding of the critical phenomenon limiting these branches of solutions is
realized  in a particular combination of the metric components the function $\Delta(r)$ defined by Eq. (\ref{Delta})
which reveals that the system of differential equations becomes singular at some
radius $r=r_c$. As far as we can see, the  limiting configurations present
no sign of an essential singularity of the metric neither signs of bifurcation into black holes or another kind of solutions.

It would be interesting to study whether the intriguing  solutions of branch 2
could exist in different versions of the tensor scalar gravity as well. As a first step
in this direction, one could check whether a self-interacting  potential (higher powers in (\ref{potential})) would lead to  branch 2-type of  solutions with  smaller values of the non-minimal parameter $\beta$.

The existence of hairy black holes in some modification of this system is also of much interest. It might be sufficient to introduce a bare cosmological constant as was done previously \cite{AnabalonEtAl2013,Minamitsuji2014} for the neutral system.
\\

\end{document}